\documentclass[preprint]{aastex631}

\usepackage{amsmath}
\usepackage{threeparttable}
\usepackage{soul}
\begin{document}

\title{Observations of Turbulence and Particle Transport at Interplanetary Shocks: Transition of Transport Regimes}

\author[0000-0003-4268-7763]{Siqi Zhao}
\affiliation{Deutsches Elektronen Synchrotron DESY, Platanenallee 6, D-15738, Zeuthen, Germany}
\affiliation{Institut für Physik und Astronomie, Universität Potsdam, D-14476, Potsdam, Germany}

\author[0000-0003-2560-8066]{Huirong Yan}
\affiliation{Deutsches Elektronen Synchrotron DESY, Platanenallee 6, D-15738, Zeuthen, Germany}
\affiliation{Institut für Physik und Astronomie, Universität Potsdam, D-14476, Potsdam, Germany}

\author[0000-0003-1778-4289]{Terry Z. Liu}
\affiliation{Department of Earth, Planetary, and Space Sciences, University of California, Los Angeles, CA 90024, USA}
 \email{siqi.zhao@desy.de}

\begin{abstract}

The transport of energetic particles is intimately related to the properties of plasma turbulence, a ubiquitous dynamic process that transfers energy across a broad range of spatial and temporal scales. However, the mechanisms governing the interactions between plasma turbulence and energetic particles remain incompletely understood. Here we present comprehensive observations from the upstream region of a quasi-perpendicular interplanetary (IP) shock on 2004 January 22, using data from four \textit{Cluster} spacecraft to investigate the interplay between turbulence dynamics and energetic particle transport. Our observations reveal a transition in energetic proton fluxes from exponential to power-law decay with increasing distance from the IP shock. This result provides possible observational evidence of a shift in transport behavior from normal diffusion to superdiffusion. This transition correlates with an increase in the time ratio from $\tau_s/\tau_{c}<1$ to $\tau_s/\tau_{c}\gg1$, where $\tau_s$ is the proton isotropization time, and $\tau_{c}$ is the turbulence correlation time. Additionally, the frequency-wavenumber distributions of magnetic energy in the power-law decay zone indicate that energetic particles excite linear Alfv\'en-like harmonic waves through gyroresonance, thereby modulating the original turbulence structure. These findings provide valuable insights for future studies on the propagation and acceleration of energetic particles in turbulent astrophysical and space plasma systems.

\end{abstract}


\section{Introduction} \label{sec:intro}

Energetic particles are ubiquitous in the highly turbulent interplanetary (IP) space, originating from diverse sources such as solar energetic particles, solar wind suprathermal particles, energetic storm particles, and anomalous cosmic rays \citep{Reames1999,Lin2005,Fisk2012,Giacalone2022}. Over the past few decades, the transport properties of these particles in turbulent environments, which are critical to their propagation and acceleration, have been extensively investigated \citep{Schlickeiser2002, Yan2008,Beresnyak2011,Lazarian2014,Zhao2021,Maiti2022}. Turbulence parameters, including anisotropy, spectral shape, and intermittency, play a significant role in shaping energetic particle transport \citep{Yan2002,Pucci2016,Yan2022,Zhao2022,Butsky2024}. Conversely, energetic particles can actively modulate turbulence dynamics via mechanisms such as gyroresonance, Laudau damping, and energization by intermittent structures \citep{Hollweg2002,Dmitruk2004, Petrosian2006,Yan2011,TenBarge2013}. This bidirectional interaction underscores the necessity for comprehensive observational studies that integrate turbulence dynamics with energetic particle behaviors, enhancing our understanding of energetic particle acceleration and energization in turbulent environments.

Normal and anomalous transport of energetic particles are found in a broad diversity of plasma systems \citep[see][]{Amato2018,Zimbardo2020,Maiti2022}. The random motion of particles is typically characterized by diffusive propagation, quantified by the diffusion coefficient $\kappa = \langle (\Delta \mathbf{x})^2 \rangle /2d't^\alpha$. Here, $\Delta \mathbf{x}$ represents the particle displacement from the initial position $\mathbf{x}'$ over the timescale $t$, with $\Delta \mathbf{x}=\mathbf{x}(t) -\mathbf{x}'$, $d'$ is the spatial dimension, and $\alpha$ is the diffusion exponent \citep{Klafter1987,Metzler2004}. For normal diffusion, the mean square displacement grows linearly with time, characterized by $\langle (\Delta \mathbf{x})^2\rangle \propto t$. In contrast, in many astrophysical and space plasma systems, particles exhibit anomalous diffusion, characterized by $\langle (\Delta \mathbf{x})^2\rangle\propto t^\alpha$ with $\alpha \neq 1$. Subdiffusion occurs when $0<\alpha<1$, indicating slow, dispersive particle transport. Conversely, superdiffusion occurs when $\alpha>1$, indicating enhanced and accelerated particle transport \citep{Metzler2004,Yan2022}. The mechanism behind superdiffusion remains unresolved, although candidate mechanisms include continuous-time random walk (CTRW)\citep{Zumofen1993}, $L\grave{e}vy$ random walk \citep{Ragot1997,Perri2007} and Richardson diffusion \citep{Richardson1926,Xu2013,Lazarian2014,Maiti2022,Hu2022}. 

Interplanetary (IP) shocks offer a natural laboratory for examining the intricate interactions between turbulence and energetic particles. Both normal diffusion and superdiffusion of energetic particles have been extensively observed in the upstream region of IP shocks \citep{Drury1983,Zimbardo2020}. This study presents comprehensive observations from the upstream region of an IP shock on 2004 January 22, using data from four \textit{Cluster} spacecraft. The findings provide the first observational evidence for the transition from normal diffusion to superdiffusion with increasing distance from the IP shock. Moreover, this study investigates how energetic protons influence three-dimensional turbulence structures by analyzing the wavenumber distributions of magnetic energy and gyroresonance conditions. The paper is organized as follows: Section 2 describes the data sets, methodology, and event selection criteria. Sections 3 and 4 focus on the analysis of energetic particles and turbulent properties in the upstream region of the IP shock, respectively. Finally, section 5 provides a summary of the key findings.

\section{Data and Methodology} 
\subsection{Data}\label{sec:data}

We used data from the \textit{Cluster} mission, which consists of four spacecraft in a tetrahedral configuration. The magnetic field data with a 0.045 s time resolution were measured by the Fluxgate Magnetometer (FGM) \citep{Balogh1997}. The proton plasma data with a 4 s time resolution were measured by the Cluster Ion Spectrometry’s Hot Ion Analyzer (CIS-HIA) \citep{Reme2001}. The electron density data from the Waves of High frequency and Sounder for Probing of Electron density by Relaxation (WHISPER) \citep{Decreau1997} were used to cross-check proton plasma data. The consistency between proton densities from CIS-HIA and electron densities from WHISPER confirms the reliability of the plasma data (Figure \ref{figure1}(c)). The energetic particle data with a 4 s time resolution were measured by the Research Adaptive Particle Imaging Detectors (RAPID), covering an energy range of 30 keV to 1500 keV (4000 keV) for protons and 20 keV to 400 keV for electrons \citep{Wilken1997}. The pitch angle distributions of energetic protons from RAPID were obtained with a $128$ s time resolution. 

\subsection{Analysis Method}\label{sec:method}

The wavenumber distributions of magnetic energy provide an approximate structure characterization of turbulence in a homogeneous and stationary state \citep{Zhao2022,Zhao2024a,Zhao2024b}. To ensure sufficient sampling, we utilized the highest-cadence data from \textit{Cluster-1}, with a time resolution of 0.045 s for the magnetic field ($\mathbf{B}$) and 4 s for the proton velocity ($\mathbf{V_p}$). The electric field data on MHD scales were calculated using $\mathbf{E} = -\mathbf{V_p} \times \mathbf{B}$. Our analysis focused on MHD-scale fluctuations, restricting the frequency range to $4/t_{win} < f_{sc} < 0.8\omega_{cp}/(2\pi)$, where $t_{win}$ is the analysis duration, $f_{sc}$ is the frequency in the spacecraft frame, and $\omega_{cp}$ is the proton gyro-frequency.

First, the time series of the magnetic field and electric field were transformed into Fourier space using the Morlet-wavelet transform \citep{Grinsted2004}. This process yielded the wavelet coefficients of the fluctuating magnetic field and electric field at each time $t$ and $f_{sc}$ in geocentric-solar-ecliptic (GSE) coordinates: $\mathbf{W_B}(t,f_{sc})=[W_{BX},W_{BY},W_{BZ}]$ and $\mathbf{W_E}(t,f_{sc})=[W_{EX},W_{EY},W_{EZ}]$. To eliminate the edge effect caused by the finite length of time series, wavelet transforms were performed over a duration twice the analysis window ($2t_{win}$), retaining only the unaffected central period for further analysis.

Second, wavevectors $\mathbf{k}(t,f_{sc})$ were calculated using the singular value decomposition (SVD) method described by \cite{Santolik2003,Zhao2021,Zhao2022,Zhao2024a,Zhao2024b}. This approach derives the wavevector by solving the linearized Faraday's law: $\mathbf{k}\times\mathbf{W_E}=2\pi f_{sc}\mathbf{W_B}$. 
To enhance reliability, the calculated $\mathbf{k}(t,f_{sc})$ were smoothed using a moving average with a window of $2^{10}$ s, as low-frequency fluctuations tend to exhibit temporal stability.

Third, the wavelet power spectra of the magnetic field at each time $t$ and $f_{sc}$ are given by
\begin{eqnarray}
P_B(t,f_{sc})=W_{BX}W_{BX}^* + W_{BY}W_{BY}^* + W_{BZ}W_{BZ}^*,
 \label{eq:1}
\end{eqnarray}
where the asterisks denote the complex conjugate. The magnetic energy at each (t,$f_{sc}$) bin is estimated as $D_B(t,f_{sc})=\delta f_{sc}P_B(t,f_{sc})$, where $\delta f_{sc}$ is the width of frequency bin. 

Fourth, the magnetic power spectra were transformed into the solar wind rest frame by correcting the Doppler shift. The frequency in the solar wind rest frame can be approximated as $f_{rest}=f_{sc}-(\mathbf{k\cdot\mathbf{V_{sw}}})$ due to the negligible spacecraft speed, where $V_{sw}$ is the solar wind velocity. This study used the representation of absolute frequencies:
\begin{align}
   (f_{rest}, \mathbf{k}) =
   \begin{cases}
       (f_{rest}, \mathbf{k}) & \text{if  } f_{rest} > 0, \\
       (-f_{rest}, -\mathbf{k}) & \text{if  } f_{rest} < 0.
   \end{cases}
\end{align}

Fifth, we constructed a set of $300\times300\times100$ bins to obtain the wavenumber-frequency distributions of magnetic power $P_B(k_\perp,k_\parallel,f_{rest})$ in the solar wind rest frame. Each bin subtends approximately the same $k_\perp$, $k_\parallel$ and $f_{rest}$ values, with widths ($dk_\perp$,$dk_\parallel$,$df_{rest}$). To ensure statistical reliability, each bin was required to contain at least 50 data points. The elongation of turbulent eddies is determined by the interactions with the local magnetic field ($\mathbf{B}_{local}$) rather than the global background field \citep{Goldreich1995,Cho2000,Zhao2022}. Therefore, $k_\parallel$ and $k_\perp$ are defined as the wavenumbers parallel and perpendicular to $\mathbf{B}_{local}$. The local magnetic field is calculated as $\mathbf{B}_{local}(t,\tau)=(\mathbf{B}(t)+\mathbf{B}(t-\tau))/2$, where the timescale is estimated as $\tau=1/f_{rest}$. 

\subsection{Event Selection Criteria}


We report a representative interplanetary (IP) shock event on 2004 January 22 that satisfies the following criteria: (1) Strong magnetic field compression and density compression, characterized by a magnetic field ratio of $B_2/B_1>3$ and a density ratio of $N_2/N_1>2$. (2) Simultaneous observations by all four \textit{Cluster} spacecraft, enabling high-quality measurements of the high-resolution magnetic field, differential particle fluxes of energetic particles, and their pitch-angle distributions. (3) A pristine upstream region, where the IP shock is embedded in the undisturbed solar wind, free from contamination by terrestrial foreshock effects caused by the near-Earth trajectory of \textit{Cluster} spacecraft. The interactions with high-energy particles reflected by the terrestrial bow shock introduce additional complexity to IP shock physics, which is beyond the scope of this study. As a supplement, we present another IP shock event observed by \textit{Cluster}-4 on 2011 February 18 in Appendix B. Both events are cataloged in the comprehensive database of IP shocks (\url{http://ipshocks.helsinki.fi}).

\section{Observations} 
\subsection{Overview}

\begin{figure*}
\centering
\includegraphics[scale=0.7]{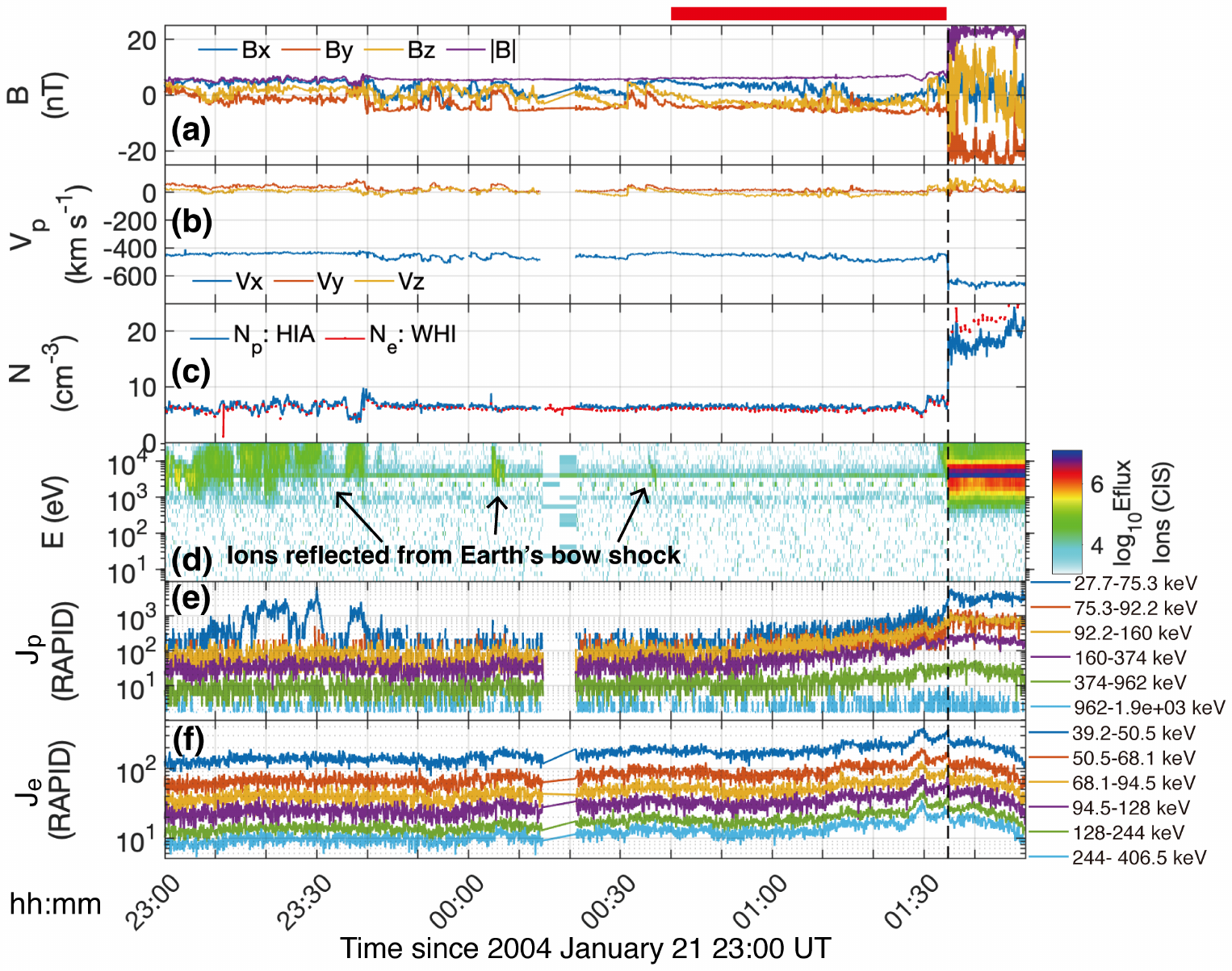}
\caption{Overview of the interplanetary shock crossing observed by \textit{Cluster}-1 from 23:00 to 01:50 UT on 2004 January 21-22. (a) Magnetic field. (b) Proton bulk velocity. (c) Proton density measured by CIS-HIA (blue) and electron density measured by WHISPER (red). (d) Energy-time spectrogram of differential energy flux (Eflux; $keVcm^{-2}s^{-1}sr^{-1}keV^{-1}$) for ions measured by CIS-HIA in high-sensitivity mode. (e,f) Differential particle flux for protons ($J_p$) and electrons ($J_e$), both in units of $cm^{-2}s^{-1}sr^{-1}keV^{-1}$, measured by RAPID. The vertical dotted line marks the time of shock passage at 01:34:44 UT. The red bar at the top of the panel (a) highlights the upstream interval analyzed in detail. All vector quantities are expressed in the GSE coordinate and in the spacecraft frame.}
\label{figure1}
\end{figure*}

Figure \ref{figure1} presents an overview of an interplanetary (IP) shock crossing, observed by \textit{Cluster-1}, from the upstream to the downstream regions. The black vertical dashed line marks the shock's arrival at 01:34:44 UT on 2004 January 22. The IP shock is characterized by the sharp increases in the magnetic field (Figure \ref{figure1}(a)), spacecraft-frame proton bulk velocity (Figure \ref{figure1}(b)), plasma density (Figure \ref{figure1}(c)), and flux intensity of particles (Figures \ref{figure1}(d-f)). At the time of the IP shock, \textit{Cluster} spacecraft were located at $[15.0,11.1,-5.2] R_E$ (Earth radius). During 23:00-00:40 UT on January 21-22, ions with energies around 10 keV and substantial magnetic field fluctuations were observed nearly simultaneously, indicating contamination by the terrestrial ion foreshock. To minimize this effect, our analysis focuses on the pristine solar wind upstream of the IP shock, highlighted by the red bar at the top of Figure \ref{figure1}.

\subsection{Transport Properties of Energetic Particles Upstream of the IP Shock}

\begin{figure*}
\centering
\includegraphics[scale=0.63]{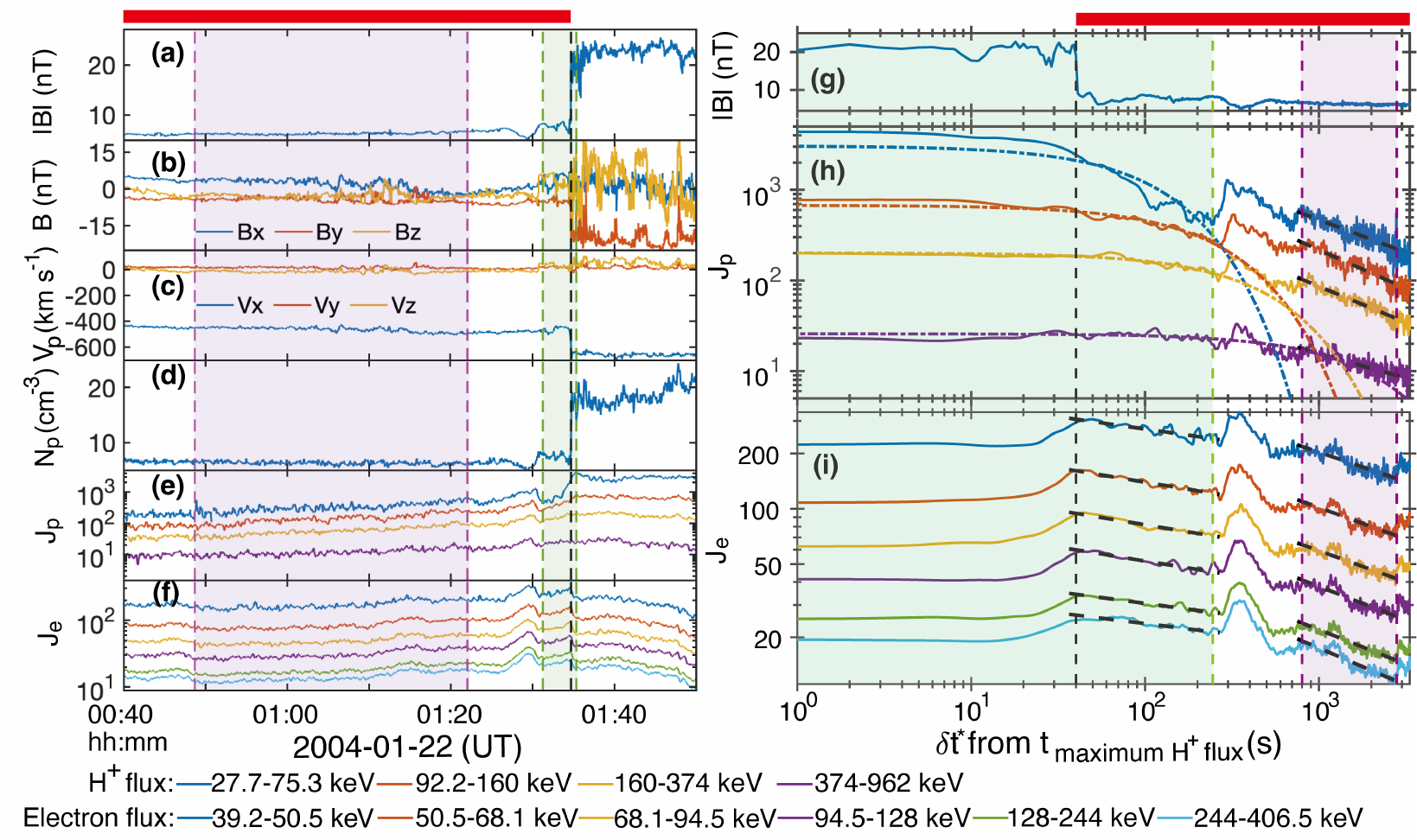}
\caption{(a,g) Magnetic field magnitude $|B|$. (b) Magnetic field components. (c) Proton bulk velocity. (d) Proton density. (e,f) Differential particle flux for protons ($J_p$) and electrons ($J_e$) (in units of $cm^{-2}s^{-1}sr^{-1}keV^{-1}$). (h,i) $J_p$ and $J_e$ are plotted in log-log axes as a function of the time interval $\delta t^*$ from the maximum proton ($H^+$) flux. The colored dashed curves denote exponential fits within the proton exponential decay zone. The black dashed lines represent power-law fits.
The black vertical dashed lines mark the shock passage at 01:34:44 UT. The purple (green) shaded regions mark the power-law (exponential) decay zone for energetic protons. The red bar at the top highlights the upstream interval of the IP shock. All vector quantities are expressed in GSE coordinates and in the spacecraft frame.}
\label{figure2}
\end{figure*}

We focus on shorter timescales to investigate the detailed transport properties of energetic particles upstream of the IP shock (Figure \ref{figure2}). All field and plasma data from all four spacecraft are interpolated to a uniform cadence of 1 s to unify the timeline. The energetic particle fluxes are averaged over 16 seconds to enhance statistical reliability. Figure \ref{figure2}(e) shows the time-intensity profiles of energetic proton flux ($J_p$), which peak at 01:35:23 UT (the second green vertical dashed line), slightly after the shock’s arrival at 01:34:44 UT (black vertical dashed line). The peak location of particle fluxes downstream behind the shock front aligns with previous studies \citep{Decker2008, Zank2015}. 

Diffusion behavior can be quantitatively described using the propagator $P(x-x',t-t')$, which defines the probability of finding a particle at position $x$ and time $t$, given its initial position $x'$ and time $t'$ \citep{Zumofen1993,Kirk1996}. Assuming particles are injected at $x'=0$ and $t'=0$, the mean square displacement simplifies to $\langle x^2(t)\rangle =\int x^2 P(x,t)dx$. For normal diffusion, energetic particles accelerated from the shock follow a Gaussian propagator. And their flux upstream of the shock exhibits exponential decay ($J\propto exp(-U_1\delta x/\kappa)$), where $U_1$ is the upstream plasma velocity in the shock rest frame, $\delta x$ is the distance from the shock, and $\kappa$ is a constant diffusion coefficient \citep{Fisk1980,Drury1983}. In contrast, for superdiffusion, the energetic particle flux upstream of the shock exhibits a power-law decay ($J\propto |\delta x|^{-\gamma}$) with a slope $0<\gamma<1$ \citep{Perri2007,Perri2008}. 

The time-intensity profiles are presented in log-log axes as a function of $\delta t^*$ in Figures \ref{figure2}(h,i), where $\delta t^*$ represents the observation time upstream of the shock relative to the maximum proton ($H^+$) flux. Assuming a constant shock speed, we approximate the upstream distance from the IP shock as $\delta x\sim V_{sh}\delta t^*$, where $V_{sh}$ is the shock normal speed calculated through mixed-mode coplanarity techniques \citep{Abraham1972} and jump conditions in \cite{Smith1988}. As the distance from the IP shock increases, the particle fluxes of both energetic protons and electrons decrease, accompanied by variations in diffusion characteristics. The upstream region of the IP shock can be roughly divided into two zones according to proton diffusion characteristics. 
\begin{enumerate}
\item \textbf{Exponential Decay Zone}: Near the shock, the flux intensity of energetic protons ($J_p$) exhibits an exponential decay, as indicated by the green-shaded region in Figure \ref{figure2} and \ref{figure3}.
\item \textbf{Power-Law Decay Zone}: Further away from the shock, the flux intensity of energetic protons ($J_p$) follows a power-law decay, as indicated by the purple-shaded region in Figure \ref{figure2} and \ref{figure3}.
\end{enumerate}
These changes in $J_p$ profiles indicate a potential transition in transport properties. Table 1 displays the shock parameters derived from upstream data in the exponential decay zone, confirming that the IP shock is quasi-perpendicular under the local upstream conditions in this region. Table 2 summarizes the basic plasma parameters in the exponential and power-law decay zones, which are used in the subsequent calculations. 

Considering that the parallel mean free path ($\lambda_\parallel$) is comparable to the size of the investigated regions (Table 2), these regions cannot be considered independent and isolated. The sharp increase in energetic particle fluxes between these two diffusion zones (from 01:22:03 to 01:30:54 UT) is likely driven by the local magnetic structure, characterized by field strength depletions and direction variations (Figure \ref{figure2}(a,b)). This local magnetic structure may also influence the transport of energetic particles in adjacent areas. For comparison, the IP shock crossing on 2011 February 18, which lacks such local magnetic structures, exhibits a smooth transition between the two transport regimes (Figure \ref{figure7} in Appendix B). A detailed investigation of how local magnetic structures affect energetic particle transport is beyond the scope of this paper, which requires further analysis of the nature of these structures (e.g., current sheets or flux tubes).


\subsubsection{Exponential Decay Zone of Energetic Protons}

Near the IP shock ($\delta t^*<245s$), the flux intensity of energetic protons decays exponentially with increasing distance upstream from the IP shock (Figure \ref{figure2}(h)). By fitting the time-intensity profiles in the interval $\delta t^*=[0,245]s$ to an exponential function, $J_p(\delta t^*)\propto exp(-\delta t^*/\Delta t_d)$, we find that higher-energy protons exhibit longer exponential diffusion timescales $\Delta t_d$. Assuming that the diffusion coefficient is independent of space and time, the diffusive skin depth can be estimated as $\Delta_d=V_{sh}\Delta t_d$. In cosmic ray physics, the diffusive skin depth ($\Delta_d$) is also referred to as the diffusion length. The component of the diffusion tensor along the shock's propagation direction can be expressed as 
\begin{eqnarray}
\kappa_{rr}=V_n\Delta_d=V_nV_{sh}\Delta t_d,
\end{eqnarray}
where $V_n$ is the shock normal component of the upstream proton bulk velocity in the normal incidence (NI) frame \citep{Giacalone2012}. We further estimated the parallel mean free path as $\lambda_{\parallel} = 3\kappa_{\parallel}/v,$
where $\kappa_{\parallel}\sim \kappa_{rr}/cos^2\theta_{Bn}$, assuming negligible perpendicular diffusion, $\theta_{Bn}$ is the shock normal angle, and $v$ is the particle velocity \citep{Zank2006,Yan2008}. Table 2 summarizes the above diffusion parameters of energetic protons. The diffusion coefficients ($k_{rr}$) in the exponential decay zone are comparable to those in the upstream region of strong IP shocks observed by ACE spacecraft \citep{Giacalone2012}.

Although the exponential diffusion upstream of the shock is consistent with the expected behavior for the diffusive shock acceleration, we remain cautious about the precise acceleration mechanism for these energetic protons, as $\lambda_\parallel$ is comparable to the size of the whole zone (on the order of $10^5 km$).


\subsubsection{Power-law Decay Zone of Energetic Protons}

When $\delta t^*>800s$, energetic proton fluxes exhibit a power-law decay, $J\propto |\delta t^*|^{-\gamma}$, as shown in Figure \ref{figure2}(h). For this event, the power-law slope $\gamma$ is 0.54-0.85 (see Table 2). Consequently, the diffusion exponent ($\alpha$) is between 1.15 and 1.46, where $\alpha=2-\gamma$ \citep{Perri2007,Perri2008}. According to the interpretation proposed by \cite{Perri2007,Perri2008}, the power-law decays of fluxes with a slope $0<\gamma<1$ indicates superdiffusive transport of energetic protons ($\langle (\Delta \mathbf{x})^2\rangle\propto t^\alpha$ with $\alpha>1$). Notably, $\alpha$ systematically increases with increasing energy, implying that higher-energy protons diffuse more rapidly and transport further from the shock. 

\subsubsection{Energetic Electron Diffusion}

In contrast to energetic protons, energetic electron fluxes ($J_e$) peak slightly before the shock's arrival (Figures \ref{figure2}(f,i)). Unlike energetic protons, $J_e$ does not exhibit an apparent exponential diffusion zone in this event. Instead, $J_e$ follows a power-law decay both near and far from the shock. The diffusion exponents, estimated as $\alpha=2-\gamma$, are consistently greater than 1, as shown in Table 3, suggesting the superdiffusion of energetic electrons. Near the shock, the diffusion exponent approaches the ballistic limit ($\alpha=2$), suggesting that electron diffusion in this region is scatter-free and proceeds significantly faster. Additionally, the diffusion exponents ($\alpha$) remain nearly consistent across different electron energies, pointing to a weaker energy dependence of electron diffusion compared to proton diffusion. 

\subsection{Pitch Angle Distributions of Energetic Protons}

\begin{figure*}
\centering
\includegraphics[scale=0.65]{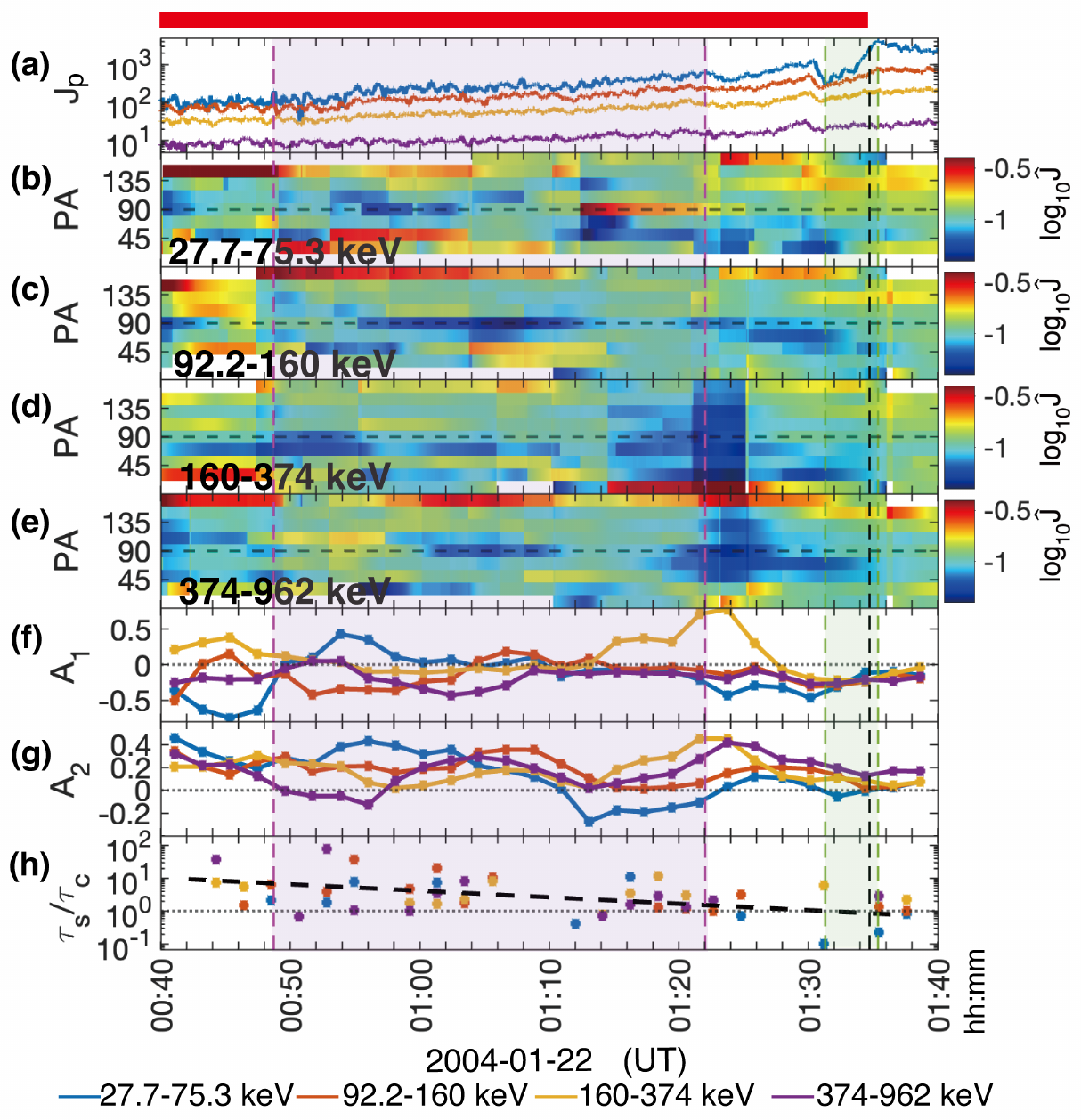}
\caption{(a) Differential particle flux for energetic protons ($J_p$) (in units of $cm^{-2}s^{-1}sr^{-1}keV^{-1}$). (b-e) Normalized pitch angle (PA) distributions for protons in the energy range 27.7-75.3 keV, 92.2-160 keV, 160-374 keV, and 374-962 keV, respectively. (f) First-order PA anisotropies ($A_1$) for energetic protons. (g) Second-order PA anisotropies ($A_2$) for energetic protons. (h) The ratio of isotropization time and correlation time ($\tau_s/\tau_{c}$). The flux intensity data were measured by the RAPID instrument onboard the four \textit{Cluster} spacecraft and moving averaged over 512 s to improve statistical reliability. The black vertical dashed line marks the shock passage at 01:34:44 UT. The purple-shaded (green-shaded) region marks the power-law (exponential) decay zone for energetic protons. The red bar at the top highlights the upstream interval of the IP shock.}
\label{figure3}
\end{figure*}

Figure \ref{figure3} presents the normalized pitch angle (PA) distributions for energetic protons in energy ranges 27.7-75.3 keV (Figure \ref{figure3}b), 92.2-160 keV (Figure \ref{figure3}c), 160-374 keV (Figure \ref{figure3}d), and 374-962 keV (Figure \ref{figure3}e). The measurements of PA distributions have some limitations, as indicated by missing data in the PA bins for $0^\circ-20^\circ$ and $160^\circ-180^\circ$. To reduce uncertainty, we resample PA distributions from nine bins into three broader categories by non-NAN averaging: parallel flux intensity $F(0^\circ-60^\circ)$, perpendicular flux intensity $F(60^\circ-120^\circ)$, and anti-parallel flux intensity $F(120^\circ-180^\circ)$. 

The first-order and second-order PA anisotropies for energetic protons are given by
\begin{eqnarray}
  A_1 =\frac{2[F(0^\circ-60^\circ)-F(120^\circ-180^\circ)]}{F(0^\circ-60^\circ)+2F(60^\circ-120^\circ)+F(120^\circ-180^\circ)},\\
  A_2 =\frac{F(0^\circ-60^\circ)-2F(60^\circ-120^\circ)+F(120^\circ-180^\circ)}{F(0^\circ-60^\circ)+2F(60^\circ-120^\circ)+F(120^\circ-180^\circ)}.
\end{eqnarray}
Figure \ref{figure3}(f) shows that the first-order PA anisotropies ($A_1$) exhibit a preference towards negative values, indicating a dominance of anti-parallel fluxes. This observation is consistent with expectations: with $B_X>0$, an anti-parallel PA distribution implies that energetic particles are transported away from the shock, which acts as the source of these particles. Figure \ref{figure3}(g) illustrates that the second-order PA anisotropies ($A_2$) are smaller in the exponential decay zone compared to the power-law decay zone, suggesting more effective pitch-angle scattering closer to the shock. The anisotropic distribution of particles excites waves through gyroresonance instability \citep[see][]{Yan2011, Bykov2013}, a process discussed further in Section 4.2. In the region nearest to the shock, $A_2$ decreases with approaching the shock, reaching its lowest values at the shock front. A similar variation in PA anisotropies is evident in another event (see Figure \ref{figure7} (i,j) in Appendix B), confirming a real trend rather than mere noise.

The isotropization time ($\tau_s$) is determined using the relation: ${\partial A_2}/{\partial t}=\tau_s^{-1}A_2$. Figure \ref{figure3}(h) shows the isotropization time relative to the correlation time ($\tau_s/\tau_{c}$), where $\tau_{c}\sim145 s$, calculated using the normalized correlation function (see Appendix A). Due to data sampling constraints, we do not distinguish between energy channels; instead, an exponential fit is applied across all data points in Figure \ref{figure3}(h). With approaching the shock, $\tau_s/\tau_{c}$ gradually decreases from being much greater than 1 in the power-law decay zone (purple-shaded region) to less than 1 in the exponential decay zone (green-shaded region). A similar trend is identified in another event (see Figure \ref{figure7} in Appendix B). 

\section{Turbulent Properties of the Magnetic Field Upstream of the IP Shock}

This section focuses on identifying interactions between turbulence and energetic protons. In the power-law decay zone (00:48:43-01:22:03 UT on 2004 January 22), magnetic field fluctuations are homogeneous and stationary (details see Appendix A), allowing for a detailed analysis of turbulence structures through wavenumber distributions of fluctuations. By contrast, fluctuations in the exponential decay zone lack homogeneity, making them unsuitable for wavenumber analysis and, therefore, excluded from further discussion.

\subsection{Wavenumber Distributions of Magnetic Energy in the Power-law Decay Zone}
\begin{figure*}
\centering
\includegraphics[scale=0.6]{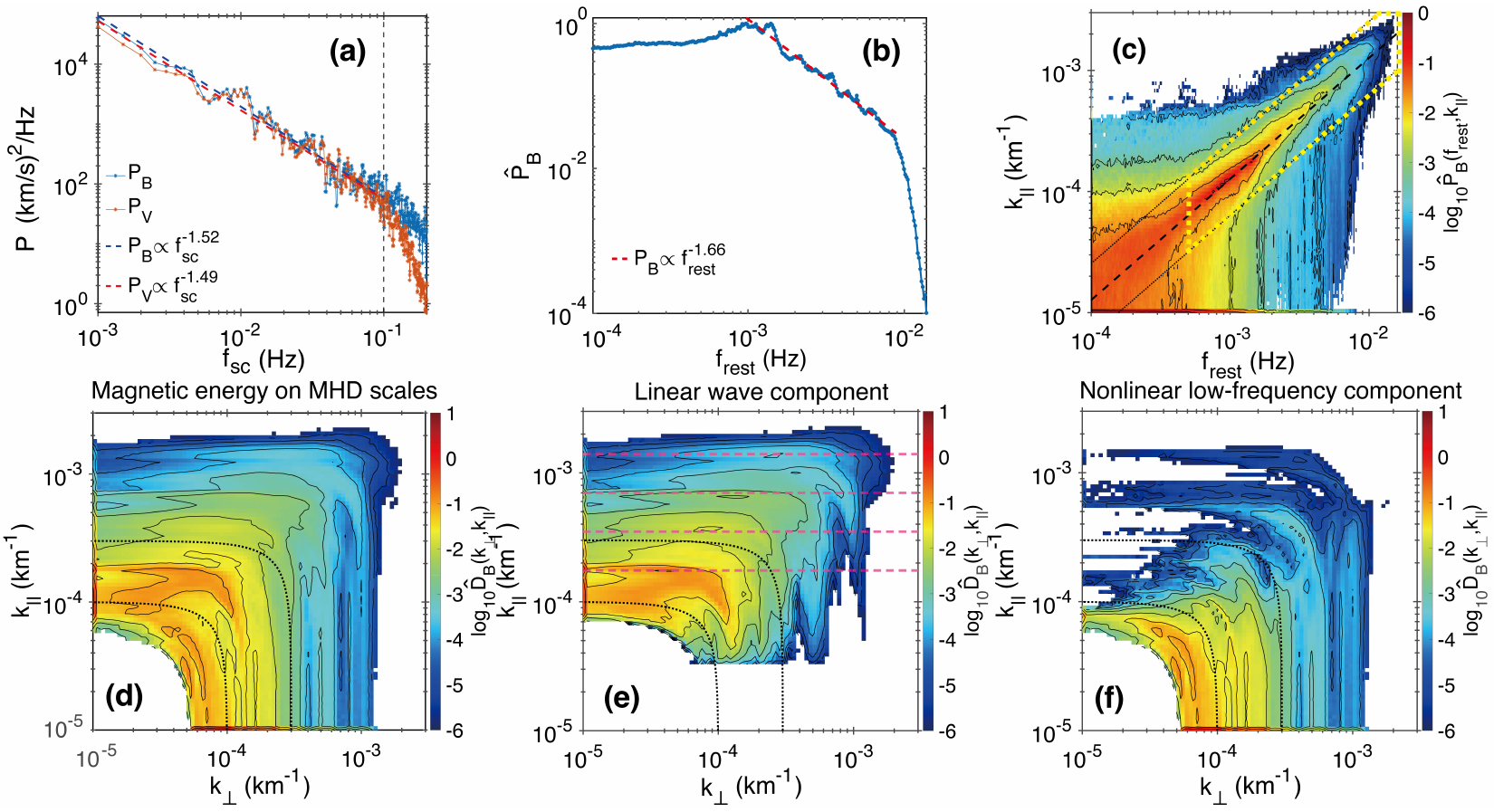}
\caption{Magnetic energy distributions in the proton power-law decay zone. (a) Power spectra density (PSD) in the spacecraft frame. The black vertical dashed line marks $f_{sc}=\omega_{cp}/(2\pi)=f_{cp}=0.1$ Hz. (b) Magnetic PSD in the solar wind rest frame, where $\hat{P}_B(f_{rest})={P}_B(f_{rest})/{P}_{B,max}$ is normalized by the maximum power across all $f_{rest}$ bins. (c) The $f_{rest}$ versus $k_{\parallel}$ distributions of magnetic power in the solar wind rest frame, where $\hat{P}_B(f_{rest},k_{\parallel})={P}_B(f_{rest},k_\parallel)/{P}_{B,max}$ is normalized by the maximum magnetic power across all $(f_{rest},k_\parallel)$ bins. The black dashed line represents the theoretical dispersion relation of Alfv\'en modes, $k_\parallel = 2\pi f_{rest}/V_A=k_A$, where $V_A$ is the Alfv\'en speed averaged over all time windows. The two black dotted lines denote $k_\parallel = 0.5k_A$ and $2k_A$. The yellow-dashed box marks the magnetic power at frequencies $f_{rest}>0.005f_{cp}$ and parallel wavenumbers $0.5k_A<k_\parallel <2k_A$. (d) Normalized wavenumber distributions of magnetic energy $\hat{D}_B(k_\perp,k_\parallel)$ on MHD scales. (e) Linear wave component of MHD-scale fluctuations: $\hat{D}_B(k_\perp,k_\parallel)$ in the frequency and wavenumber range highlighted by the yellow-dashed box in panel (c). Magnetic energy peaks at $k_\parallel=k^*=1.75\times10^{-4} km^{-1}$, with the pink dashed lines marking $k^*$, $2k^*$, $4k^*$, and $8k^*$. (f) Nonlinear low-frequency component of MHD-scale fluctuations: $\hat{D}_B(k_\perp,k_\parallel)$ in the frequency and wavenumber range outside the yellow-dashed box in panel (c). For comparison, $\hat{D}_B(k_\perp,k_\parallel)$ in panels (d-f) are normalized by the same constant and displayed using the same color map. The black dotted curves in Panel (d-f) mark $k=\sqrt{k_\parallel^2+k_\perp^2}=10^{-4} km^{-1}$ and $3\times10^{-4} km^{-1}$.}
\label{figure4}
\end{figure*}

Figure \ref{figure4}(a) shows the trace power spectral density (PSD) for both the magnetic field and proton velocity, calculated using fast Fourier transform (FFT) with three-point smoothing. The trace power is defined as $P_B=P_{BX}+P_{BY}+P_{BZ}$ and $P_V=P_{VX}+P_{VY}+P_{VZ}$, where the magnetic field is expressed in Alfv\'en (velocity) units ($\delta\mathbf{B}/\sqrt{\mu_0\rho_0}$), $\delta$ indicates the use of a fluctuating quantity, $\mu_0$ is the vacuum permeability, and $\rho_0$ is the mean proton mass density \citep{Bruno2013}. The alignment between $P_B$ and $P_V$ indicates a high Alfv\'enicity of fluctuations. Moreover, the power-law index is approximately $-1.5$ for $f_{sc}<\omega_{cp}/(2\pi)=f_{cp}$, implying that fluctuations are in a well-developed state.

Using the method described in Section \ref{sec:method}, we further calculated the wavenumber-frequency distributions of magnetic power $P_B(k_\perp,k_\parallel,f_{rest})$ in the solar wind rest frame. To ensure sufficient sampling, we applied a moving time window with a 20-minute length and a 1-minute step and then integrated data points across all time windows. The relative amplitudes of the magnetic field, given by $\delta B_{rms}/|B_0|=\sqrt{|\mathbf{B}-\langle\mathbf{B}\rangle_{twin}|^2}/\langle\mathbf{B}\rangle_{twin}\sim s\pm \sigma \sim 0.31\pm0.20$, suggests the fluctuating magnetic field with a low turbulent Alfv\'en Mach number. Here, $s$ and $\sigma$ denote the average value and standard deviation.

The one-dimensional reduced magnetic power in the solar wind rest frame is given by 
\begin{align}
P_B(f_{rest}) =\int_{0}^{\infty}\int_{0}^{\infty}P_B(k_\perp,k_\parallel,f_{rest})dk_\perp dk_\parallel 
\sim \sum_{k_{\perp,min}}^{k_{\perp,max}}\sum_{k_{\parallel,min}}^{k_{\parallel,max}}P_B(k_\perp,k_\parallel,f_{rest})dk_\perp dk_\parallel. 
\end{align}
The wavenumber-frequency distribution of magnetic power is given by 
\begin{eqnarray}
P_B(k_\parallel,f_{rest})=\int_{0}^{\infty}P_B(k_\perp,k_\parallel,f_{rest})dk_\perp
\sim \sum_{k_{\perp,min}}^{k_{\perp,max}}P_B(k_\perp,k_\parallel,f_{rest})dk_\perp.
\end{eqnarray}
The wavenumber distribution of magnetic energy is given by 
\begin{eqnarray}
    D_B(k_\perp,k_\parallel) = \int_{0}^{\infty}P_B(k_\perp,k_\parallel,f_{rest})df_{rest} \sim \sum_{f_{rest,min}}^{f_{rest,max}}P_B(k_\perp,k_\parallel,f_{rest})df_{rest}.
\end{eqnarray}
Here, $k_{\parallel,min}=k_{\perp,min}=10^{-5}km^{-1}$, $k_{\parallel,max}=k_{\perp,max}=10^{-2}km^{-1}$, $f_{rest,min}=10^{-3}f_{cp}=10^{-4}$ Hz, and $f_{rest,max}=0.2f_{cp}=2\times10^{-2}$ Hz. We note that the presence of magnetic energy at the smallest wavenumber bins in Figure \ref{figure4}(c-f) corresponds to magnetic energy at larger length scales. Due to the limitations of spacecraft separation (approximately 200 km in this event), energy distributions at these larger scales are less reliable and therefore are not discussed in this study. 

Figure \ref{figure4}(b) shows that the normalized magnetic power ($\hat{P}_B(f_{rest})$) remains roughly constant at $f_{rest}<0.005f_{cp}$, whereas it follows a Kolmogorov-like spectrum: $\hat{P}_B\propto f_{rest}^{-1.66}$ at $0.01f_{cp}<f_{rest}<0.1f_{cp}$, where the gyro-frequency $f_{cp}=\omega_{cp}/(2\pi)=0.1$ Hz. Figure \ref{figure4}(c) presents the normalized magnetic power $\hat{P}_B(f_{rest},k_\parallel)$, which includes a linear wave component and nonlinear low-frequency fluctuations. We distinguish these two parts of fluctuations by limiting the frequency and parallel wavenumber ranges as follows: 
\begin{enumerate}
\item \textbf{Linear wave component}: For $f_{rest}>0.005f_{cp}$, most magnetic power is concentrated within the parallel wavenumber range $0.5k_A<k_\parallel<2k_A$ (inside the yellow-dashed box in Figure \ref{figure4}(c)). The Alfv\'en parallel wavenumber $k_A$ is estimated using the theoretical dispersion relations of Alfv\'en modes ($k_A =2\pi f_{rest}/V_A$), where $V_A=B_0/\sqrt{4\pi\rho_0}$ is the Alfv\'en velocity, and $B_0\sim6$ nT is the mean magnetic field over all time windows.

\item \textbf{Nonlinear low-frequency fluctuations}: For $f_{rest}<0.005f_{cp}$, most magnetic power significantly deviates from the dispersion relations of Alfv\'en modes (falling outside the yellow-dashed box in Figure \ref{figure4}(c)). 
\end{enumerate}
This method of distinguishing fluctuations has limitations, as single-frequency waves and nonlinear low-frequency fluctuations are inherently intertwined. Therefore, only approximate distinctions can be made when analyzing the turbulence properties of each component. 

Figure \ref{figure4}(e) shows the linear wave component of the fluctuations, which closely aligns with the dispersion relations of Alfv\'en modes. Overall, the wavenumber distribution of magnetic energy ($\hat{D}_B(k_\perp,k_\parallel)$) is concentrated within a propagation angle range of $\theta<45^\circ$. Additionally, $\hat{D}_B(k_\perp,k_\parallel)$ primarily decreases as $k_\parallel$ increases, whereas remaining relatively stable along $k_\perp$. Notable peaks are observed at $k_\parallel=k^*=1.75\times10^{-4}km^{-1}$ and its integer multiples, as marked by the pink dashed lines.

Figure \ref{figure4}(f) displays the wavenumber distribution of the nonlinear low-frequency fluctuations. The magnetic energy $\hat{D}_B(k_\perp,k_\parallel)$ is predominantly distributed along the $k_\perp$ direction, suggesting a quasi-two-dimensional (quasi-2D) cascade. The parallel energy spectrum of these nonlinear low-frequency fluctuations shows that $\hat{D}(k_\parallel)$ remains constant with increasing $k_\parallel$ at small parallel wavenumbers ($k_\parallel<6\times10^{-5}km^{-1}$; Figure \ref{figure8} in Appendix C). Additionally, the perpendicular energy spectrum follows a scaling of $\hat{D}(k_\perp)\propto k_\perp^{-2}$, further supporting the interpretation that these nonlinear low-frequency fluctuations exist in a quasi-2D weak turbulence state \citep{GALTIER2000,Galtier2003,Makwana2020,Zhao2024a}.

\subsection{Wave Amplification through Gyroresonance Instability}

MHD perturbations can grow via gyroresonance instability when their gyro-frequencies resonate with the fluctuating field \citep[see][]{Yan2011,Bykov2013}. We estimate the parallel resonance wavenumber ($k_{i\parallel}$) based on the cyclotron resonant conditions of wave-particle interactions, given by \citep{Jokipii1966,Kennel1966}
\begin{eqnarray}
    \omega_i - k_{i\parallel}v_{i\parallel}= \pm n\omega_{cp}. \qquad  (n = 1,2,3,...) 
\end{eqnarray}
Here, $v_{i\parallel}=\mu v_{i}$ is the parallel velocity, $\mu$ is the cosine of the pitch angle (PA), $v_{i}=\sqrt{2E_i/m_p}$ is the particle velocity, $E_i$ is the particle energy, the subscript $i$ denotes energy channels, $m_p$ is the proton mass, and $\omega_{cp}$ is the proton gyro-frequency. For energetic protons, the wave frequency $\omega_i \ll k_{i\parallel}v_{i\parallel}$, allowing the parallel resonance wavenumber to be approximated as $k_{i\parallel}=|n\omega_{cp}/v_{i\parallel}|\geq| n\omega_{cp}/v_{i}|=nk_{i\parallel,min}$. Table 4 shows that the range of $k_{i\parallel,min}$ matches closely with the wavenumbers where $\hat{D}_B(k_\perp,k_\parallel)$ peaks ($k_\parallel\sim k^*=1.75\times10^{-4}km^{-1}$). This alignment strongly suggests that the linear wave component of magnetic energy is likely generated through gyroresonance with energetic protons. 

Notably, $\hat{D}_B(k_\perp,k_\parallel)$ also exhibits local peaks at integer multiples of $k^*$ (i.e., $2k^*$, $4k^*$, and $8k^*$), marked by pink dashed lines in Figure \ref{figure4}(e) and Figure \ref{figure9} in Appendix D. This pattern indicates the presence of higher-order harmonics of the fundamental Alfv\'en wave. We deduce that the nonlinear terms likely play a significant role in generating these harmonics, with second-order nonlinear terms ($u^2$) significantly outweighing higher-order nonlinear terms (e.g., $u^3$, $u^4$, etc.). The $u^2$ terms likely originate from the expansion of $\frac{\partial^2u}{\partial t^2}-v'^2\frac{\partial^2u}{\partial x^2}\sim a_1u^2$, or from the expansion of $\frac{d u}{dt}=\frac{\partial u}{\partial t} + (u\cdot \nabla) u$, where the fixed coefficient $v'$ represents the wave propagation speed, and $a_1$ is the nonlinear coefficient \citep{landau1987}. Assuming that the injection wave is a single-frequency cosine wave ($u\sim u_0 exp[i(kx-\omega t)]$), the $u^2$ terms generate the second harmonic, which subsequently leads to the fourth harmonic through the same mechanism, followed by the eighth and sixteenth harmonics in succession. Here, $u_0$ is the wave amplitude, and $\omega$ is the wave frequency. Similar waves and their harmonics have been observed at the magnetospheric cusp \citep{Cargill2005}, where they are likely generated by sheared bulk flow. However, this mechanism does not apply to our observations, due to the absence of a significant shear flow in the upstream region of the IP shock.

The gyroresonance between energetic particles and weakly turbulent fluctuations can be described by the standard quasi-linear theory (QLT) \citep{Jokipii1966,Earl1974,Schlickeiser1993}. Under QLT, the minimum parallel mean free path is calculated as $\lambda_{i\parallel,min} \sim r_{ci}B_0^2/\delta B^2(k_{i\parallel,min})$, where the gyro-radius is defined as $r_{ci}=v_i/\omega_{cp}$, and $\delta B^2(k_{i\parallel,min})$ is
the magnetic energy at the minimum parallel resonance wavenumber. $\lambda_{i\parallel,min}$ is on the order of $10^5$ km, significantly larger than the turbulent injection scale. This implies that energetic protons undergo little scattering during their propagation. This finding is consistent with the large pitch angle anisotropy observed in the power-law decay zone in Figure \ref{figure3}.

The growth rate of parallel-propagating transverse modes by gyroresonance instability can be approximated as follows \citep{Lebiga2018}
\begin{eqnarray}
    \Gamma_{gr}(k)\sim \frac{5\pi}{8}\frac{\xi-1}{\xi+1}(\frac{c}{V_A})(\frac{N_{nt}}{N_0})(kr_{0})^{\xi-1}\omega_{cp}A_2,
\end{eqnarray}
where $c$ is the light speed, $N_0$ is the mean proton density, $dN/dp\propto p^{-\xi}$, and $p$ is the kinetic momentum. The parameter $\xi$ is determined by the relationship $\xi=2\eta-1$, where $dN/dE\propto E^{-\eta}$. In the power-law decay zone, $\eta\sim1.9-2.3$ for $E\geq75.3$ keV (Figure \ref{figure10} in Appendix E), suggesting that protons with $E\geq75.3$ keV are nonthermal. The parameter $r_0$ is the minimum gyro-radius of nonthermal protons (at $E$ = 75.3 keV). The density of nonthermal protons is estimated as $N_{nt}=4\pi\int_{v_1}^{v_5} v^2f(v)dv$, where $v_1$ and $v_5$ correspond to the velocities of 75.3 keV and 962 keV protons, respectively. The velocity distribution is calculated as $f(v)[s^3/m^6]=\frac{\gamma'^3 m_pJ_p}{v^2}=6.25\times10^{19}\gamma'^3\frac{m_p[kg]J_p[cm^{-2}s^{-1}sr^{-1}keV^{-1}]}{v^2[m/s]}$, and $\gamma'$ is the relativistic factor \citep{Wang2021}. The parameter $A_2$ is the second-order pitch angle anisotropy. 

Table 4 summarizes the growth rates at the minimum parallel resonance wavenumbers ($\Gamma_{gr}(k_{i\parallel,min})$). In the power-law decay zone, $\overline{A_2}=0.145\pm0.035$ represents the average over energy channels, with the uncertainty expressed as a standard deviation (Figure 3(g)). The growth rates ($\Gamma_{gr}\sim 0.003-0.07s^{-1}$) substantially exceed the damping rates, including both the nonlinear Landau damping and turbulence damping, which are on the order of $10^{-4}s^{-1}$, at similar parallel wavenumbers. This finding indicates that these wave fluctuations are rapidly excited locally, likely driven by the substantial presence of anisotropic energetic protons, and decay at a much slower rate. In addition to damping, the feedback from the scattering by the instability-induced waves is insufficient to quench the anisotropy, allowing continued wave growth. It is noteworthy that $\eta$ decreases farther away from the shock (Figure \ref{figure10}), reflecting the hardening of the particle energy spectrum. It aligns with our expectations since higher-energy particles are transported faster than lower-energy ones in the superdiffusion regime.


\section{Summary}

Understanding particle transport is essential to uncovering the mechanisms underlying particle acceleration in turbulent environments. This study presents \textit{Cluster} observations from the upstream region of an interplanetary (IP) shock on 2004 January 22, to investigate the interactions between turbulence dynamics and energetic particle transport. we provide the first observational evidence for a transition from normal diffusion to a superdiffusion regime.

The main conclusions of this study are listed as follows:
\begin{enumerate}
\item Energetic proton fluxes exhibit a transition from exponential to power-law decay with increasing distance from the IP shock. This shift in decay profiles suggests a change in transport properties, evolving from normal diffusion ($\langle (\Delta \mathbf{x})^2\rangle \propto t$) to superdiffusion ($\langle (\Delta \mathbf{x})^2\rangle \propto t^\alpha$, with $\alpha>1$).

\item The isotropization time of energetic protons increases with increasing distance from the IP shock, transitioning from $\tau_s/\tau_{c}<1$ in the exponential decay zone to $\tau_s/\tau_{c}\gg 1$ in the power-law decay zone. This suggests that energetic protons progressively exhibit superdiffusive transport as the relative scale $\tau_s/\tau_{c}$ grows.

\item In the power-law decay zone of energetic protons, the wavenumber distributions of magnetic energy consist of linear Alfv\'en-like components and nonlinear low-frequency fluctuations. Energetic protons excite linear Alfv\'en-like harmonic waves through gyroresonance, thereby modulating the original turbulence structure. 

\end{enumerate}

We would like to thank the members of the \textit{Cluster} spacecraft team and NASA’s Coordinated Data Analysis Web. The \textit{Cluster} data are available at \url{https://cdaweb.gsfc.nasa.gov}. This paper uses data from the Heliospheric Shock Database (\url{http://ipshocks.helsinki.fi}), generated and maintained at the University of Helsinki. Data analysis was performed using the IRFU-MATLAB analysis package \citep{Khotyaintsev2024} available at \url{https://github.com/irfu/irfu-matlab}. We sincerely thank the reviewers for their valuable comments and suggestions, which have greatly improved the quality of this manuscript.

\appendix

\section{Examination of the Turbulence State}
\begin{figure*}
\centering
\includegraphics[scale=0.26]{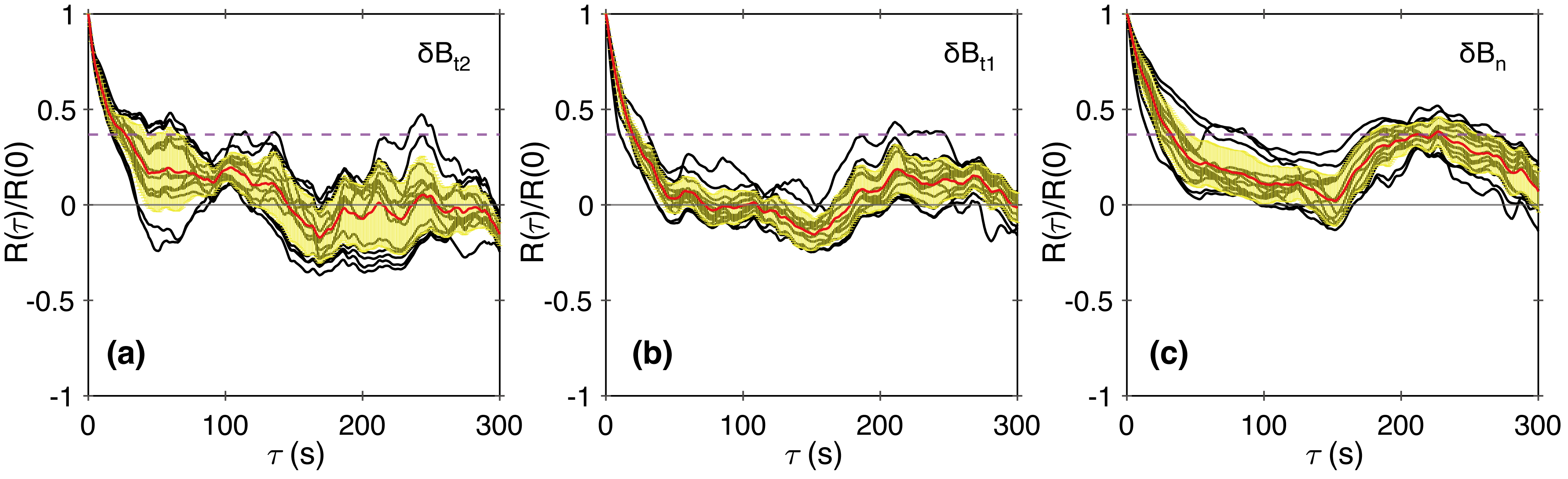}
\caption{Normalized correlation function $R(\tau)/R(0)$ as a function of timescale $\tau$ for the magnetic field in the power-law decay zone of energetic protons. The magnetic field is shown in shock-aligned coordinates, defined by $\hat{\mathbf{n}}$, $\hat{\mathbf{t}}_2=\frac{\hat{\mathbf{n}}\times\mathbf{\overline{B}}}{|\hat{\mathbf{n}}\times\mathbf{\overline{B}}|}$, and $\hat{\mathbf{t}}_1 = \hat{\mathbf{t}}_2\times\hat{\mathbf{n}}$, where $\hat{\mathbf{n}}$ represents the shock normal direction, and $\mathbf{\overline{B}}$ is the average magnetic field over 00:48:43-01:22:03 UT on 2004 January 22. The red curves represent average values ($s$) of $R(\tau)/R(0)$ over the corresponding time windows. The yellow-shaded regions represent $[s-\sigma,s+\sigma]$, where $\sigma$ is the standard deviation. The horizontal lines mark $R(\tau)/R(0)=0$ and $1/e$, where $e$ is natural constant.}
\label{figure5}
\end{figure*}

To examine the turbulent state, we calculated the correlation time ($\tau_{c}$), defined as $\tau_{c}=\int_0^\infty d\tau R(\tau)/R(0)$ \citep{Matthaeus1982}. The two-point correlation function of the magnetic field is given by $R(\tau)=\langle\delta B(t)\delta B(t+\tau)\rangle$, where $\delta B= B(t)-\langle B(t)\rangle$, $\tau$ is the timescale, and $\langle...\rangle$ is a time average. Figure \ref{figure5} presents $R(\tau)/R(0)$ for the three components of the magnetic field in shock-aligned coordinates within the proton power-law decay zone, calculated using a moving time window with a 20-minute length. The correlation time $\tau_{c}\sim[50,150]s$, significantly shorter than the time window length, indicates that the fluctuations are in a stationary state \citep{Matthaeus1982}. Additionally, the consistent $R(\tau)/R(0)$ profiles across all windows suggest that the starting time of the moving window has minimal effect on $R(\tau)/R(0)$, supporting the homogeneity of the fluctuations.

\begin{figure*}
\centering
\includegraphics[scale=0.62]{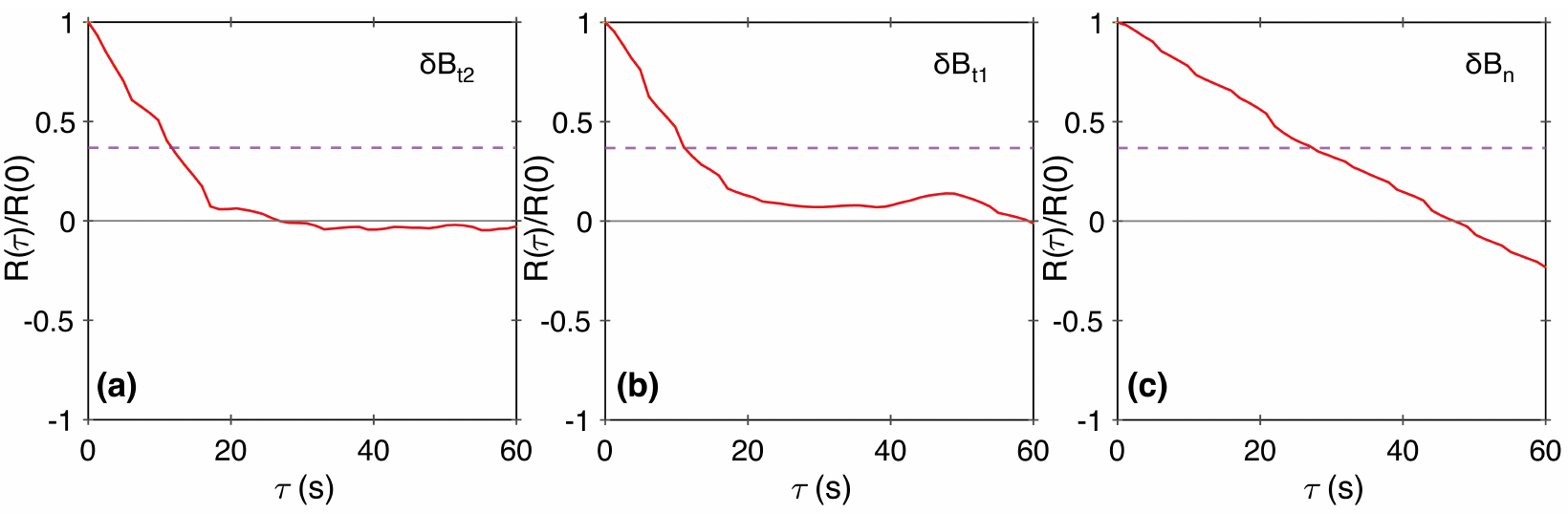}
\caption{Normalized correlation function $R(\tau)/R(0)$ as a function of timescale $\tau$ (red curves) for the magnetic field in the exponential decay zone of energetic protons. The magnetic field is shown in shock-aligned coordinates, defined by $\hat{\mathbf{n}}$, $\hat{\mathbf{t}}_2=\frac{\hat{\mathbf{n}}\times\mathbf{\overline{B}}}{|\hat{\mathbf{n}}\times\mathbf{\overline{B}}|}$, and $\hat{\mathbf{t}}_1 = \hat{\mathbf{t}}_2\times\hat{\mathbf{n}}$, where $\hat{\mathbf{n}}$ represents the shock normal direction, and $\mathbf{\overline{B}}$ is the average magnetic field over 01:30:54-01:34:29 UT. The horizontal lines mark $R(\tau)/R(0)=0$ and $1/e$.}
\label{figure6}
\end{figure*}

Figure \ref{figure6} presents $R(\tau)/R(0)$ in the proton exponential decay zone. Due to the short duration of this zone, $R(\tau)/R(0)$ was calculated using a single time window spanning 01:30:30–01:34:30 UT. The correlation time for the $\delta B_{t1}$ and $\delta B_{t2}$ components is $\tau_{c}\sim[27,40]s$, significantly shorter than the 4-minute time window. Thus, the fluctuations can be approximated as stationary in shock tangential directions. However, for the $\delta B_{n}$ component, a steady value of $R(\tau)/R(0)$ was not observed, and thus $\tau_{c}$ cannot be determined for this direction. 

Considering that a longer time window length may result in longer correlation times, part of the difference in $\tau_{c}$ can be attributed to the difference in time window length \citep{Zhao2022}. Nevertheless, in this event, the correlation times determined in the power-law and exponential decay zone are of the same order. The time window in the power-law decay zone is five times as long as that in the exponential decay zone. Therefore, the isotropization time was approximately normalized using a consistent $\tau_c\sim 145s$ in Figure \ref{figure3}. The uncertainties mentioned above do not affect our main conclusions.

\section{Universal Analysis of the Correlation Between the Proton Diffusion Regime and the Relative Scale of Proton Isotropization Time to Turbulence Correlation Time}

\begin{figure*}
\centering
\includegraphics[scale=0.65]{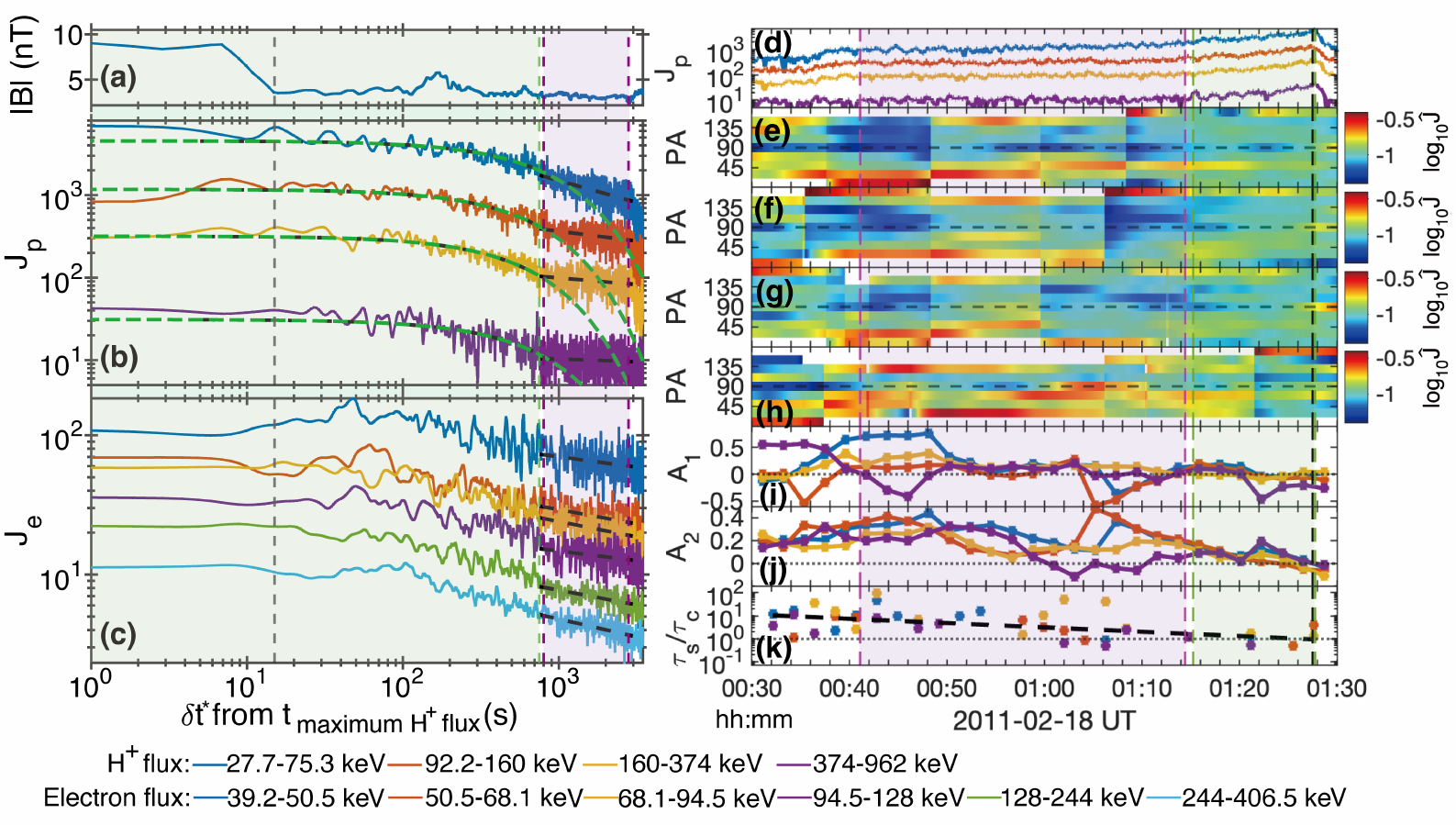}
\caption{(a) Magnetic field magnitude. (b,c) Differential particle flux $J_p$ and $J_e$ (in units of $cm^{-2}s^{-1}sr^{-1}keV^{-1}$). Panels (a-c) are plotted in log-log axes as a function of the time interval $\delta t^*$ from the maximum proton flux. The green (black) dashed curves represent the exponential (power-law) fits. (d) $J_p$. (e-h) Normalized pitch angle (PA) distributions for protons in the energy ranges 27.7-75.3 keV, 92.2-160 keV, 160-374 keV, and 374-962 keV, respectively. (i) First-order PA anisotropies ($A_1$) for energetic protons. (j) Second-order PA anisotropies ($A_2$) for energetic protons. (k) The ratio of proton isotropization time and turbulence correlation time ($\tau_s/\tau_{c}$). The flux intensity data were moving averaged over 512 s to improve statistical reliability. The black vertical dashed lines mark the shock passage. The purple (green) shaded regions mark the power-law (exponential) decay zone for energetic protons.}
\label{figure7}
\end{figure*}

Figure \ref{figure7} presents the observations of an interplanetary (IP) shock crossing at 01:27:30 UT on 2011 February 18, observed by \textit{Cluster}-4. In Figure \ref{figure7}(b), energetic proton fluxes transition from exponential decay to power-law ($\gamma<1$) decay with increasing distance from the IP shock, indicating a shift from normal diffusion ($\langle (\Delta \mathbf{x})^2\rangle \propto t$) to superdiffusion ($\langle (\Delta \mathbf{x})^2\rangle \propto t^\alpha$, with $\alpha>1$). Additionally, this transition in diffusion regimes is accompanied by an increase in proton isotropization time, with $\tau_s/\tau_{c}$ rising from less than 1 to much greater than 1, as illustrated in Figure \ref{figure7}(k).

\section{One-dimensional Reduced Wavenumber Distributions of Magnetic Energy}

Figure \ref{figure8} shows the one-dimensional reduced wavenumber distributions in the power-law decay zone. The one-dimensional reduced parallel wavenumber distributions of magnetic energy on MHD scales are given by 
\begin{eqnarray}
    D_B(k_\parallel) = \int_0^{\infty}\int_{0}^{\infty}P_B(k_\perp,k_\parallel,f_{rest})dk_\perp df_{rest}
     \sim \sum_{k_{\perp,min}}^{k_{\perp,max}} \sum_{f_{rest,min}}^{f_{rest,max}}P_B(k_\perp,k_\parallel,f_{rest})dk_\perp df_{rest},
\end{eqnarray}
where $k_{\perp,min}=10^{-5}km^{-1}$, $k_{\perp,max}=10^{-2}km^{-1}$,
$f_{rest,min}=10^{-4}$ Hz, and $f_{rest,max}=2\times10^{-2}$ Hz. The parallel wavenumber distributions of the linear wave components are given by
\begin{eqnarray}
    D_B(k_\parallel) \sim \sum_{\frac{1}{2}k_{A}}^{2k_{A}} \sum_{f_{rest,min}}^{f_{rest,max}}P_B(k_\perp,k_\parallel,f_{rest})dk_\perp df_{rest},
\end{eqnarray}
where the Alfv\'en wavenumber is calculated by $k_{A}=2\pi f_{rest}/V_A$, $f_{rest,min}=0.005$ Hz, and $f_{rest,max}=2\times10^{-2}$ Hz. The nonlinear low-frequency part of magnetic energy is the remaining magnetic energy after removing the linear wave components. 

Similarly, the one-dimensional reduced perpendicular wavenumber distributions of magnetic energy on MHD scales are given by 
\begin{eqnarray}
    D_B(k_\perp) = \int_0^{\infty}\int_{0}^{\infty}P_B(k_\perp,k_\parallel,f_{rest})dk_\parallel df_{rest}
 \sim \sum_{k_{\parallel,min}}^{k_{\parallel,max}} \sum_{f_{rest,min}}^{f_{rest,max}}P_B(k_\perp,k_\parallel,f_{rest})dk_\parallel df_{rest},
\end{eqnarray}
where $k_{\parallel,min}=10^{-5}km^{-1}$, $k_{\parallel,max}=10^{-2}km^{-1}$,
$f_{rest,min}=10^{-4}$ Hz, and $f_{rest,max}=2\times10^{-2}$ Hz. The perpendicular wavenumber distributions of the linear wave components are given by
\begin{eqnarray}
    D_B(k_\perp) \sim \sum_{\frac{1}{2}k_{A}}^{2k_{A}} \sum_{f_{rest,min}}^{f_{rest,max}}P_B(k_\perp,k_\parallel,f_{rest})dk_\parallel df_{rest},
\end{eqnarray}
where the Alfv\'en wavenumber is calculated by $k_{A}=2\pi f_{rest}/V_A$, $f_{rest,min}=0.005$ Hz, and $f_{rest,max}=2\times10^{-2}$ Hz. The nonlinear low-frequency part of magnetic energy is the remaining magnetic energy after removing the linear wave components. 

\section{Examination of harmonics of the fundamental Alfv\'en wave}

In Figure \ref{figure4}(e), two-dimensional (2D) magnetic energy $D_B(k_\parallel,k_\perp)$ exhibits local peaks at multiples of $k^*$, suggesting the presence of harmonics of the fundamental Alfv\'en wave in the linear wave component of fluctuations. To further investigate these harmonics, we calculated $\hat{D}_B(k_\parallel)$ by integrating $D_B(k_\parallel,k_\perp)$ over $k_\perp$. In Figure \ref{figure9}, blue, red, yellow, purple, and green curves correspond to the ranges $k_\perp$=[$1\times 10^{-5}$,$\infty$] $km^{-1}$, [$3\times 10^{-5}$,$\infty$] $km^{-1}$, [$6\times 10^{-5}$,$\infty$] $km^{-1}$,[$1.2\times 10^{-4}$,$\infty$] $km^{-1}$, and [$2.4\times 10^{-4}$,$\infty$] $km^{-1}$, respectively. This one-dimensional (1D) magnetic energy $\hat{D}_B(k_\parallel)$ peaks at a slightly smaller $k_\parallel$ than $k^*=1.75\times10^{-4} km^{-1}$, where 2D magnetic energy $D_B(k_\parallel,k_\perp)$ reaches its maximum in Figure \ref{figure4}(e). This shift is likely due to the incomplete decomposition of linear wave components and nonlinear low-frequency fluctuations. Nevertheless, $\hat{D}_B(k_\parallel)$ still exhibits local peaks at integer multiples of $k^*$. Furthermore, higher-order harmonics appear only at $k_\parallel=2k^*$, $4k^*$, and $8k^*$ (pink dashed lines), whereas harmonics are not significant at $k_\parallel=3k^*$, $5k^*$, $6k^*$, $7k^*$, and $9k^*$ (light blue dashed lines).

\section{Energy spectra of Energetic protons}

Figure \ref{figure10} shows the variation in the energy spectra of energetic protons upstream of the IP shock observed on 2004 January 22. The energy spectra follow a power-law scaling of $dN/dE\propto E^{-\eta}$. The partial number density is calculated from the differential particle flux ($J_p$) using the relation: 
\begin{eqnarray}
    \frac{dN[cm^{-3}]}{dE[keV^{-1}]} = 4\pi C_0 (\frac{m_p[kg]}{2})^{1/2}\frac{J_p[cm^{-2}s^{-1}sr^{-1}keV^{-1}]}{(E[keV])^{1/2}}.
\end{eqnarray}
Here, $C_0=1.60218\times10^{-18}$ is a unit conversion factor, and $E$ is the energy corresponding to specific energy channels. Figure \ref{figure10}(b) shows that the index $\eta$ increases with approaching the IP shock. 

\begin{table*}[ht]
\label{table1}
\tablenum{1}
\caption{Parameters of the IP Shock Crossing at 01:34:44 UT on 2004 January 22 for the Exponential Decay Zone}
\centering
\begin{tabular}{|c|c|}
\hline
Upstream interval (UT)  & 01:30:54-01:34:29 \\\hline
Downstream interval (UT)& 01:38:00-01:46:00 \\\hline
Shock normal direction $\hat{\mathbf{n}}$ & [-0.98,0.16,0.13] \\ \hline
$\theta_{Bn}$ ($^\circ$)  & 57\\\hline
Magnetic compression ratio $B_2/B_1$ & 3.1\\\hline
Density compression ratio $N_2/N_1$ & 2.4\\ \hline
Shock speed $V_{sh}$ ($km s^{-1}$)  & 728 \\ \hline
Upstream NI frame velocity $V_{NIF}$ ($km s^{-1}$) &  [-724.9,61.6,67.0]\\\hline
$V_n$ in NI frame ($km s^{-1}$) & -268.3 \\ \hline
Alfv\'en Mach number & 4.8 \\ \hline
\end{tabular}
\begin{tablenotes}
  \item 1. The subscripts 1 and 2 represent the upstream and downstream of the IP shock.
  \item 2. All vectors are present in GSE coordinates.
\end{tablenotes}
\end{table*}

\begin{table*}[ht]
\label{table2}
\tablenum{2}
\caption{Diffusion Parameters of Energetic Protons Upstream of the IP Shock on 2004 January 22}
\centering
\begin{tabular}{|c||c|c|c|c||c|c|}
\hline
\textbf{Proton} & \multicolumn{4}{c||}{\textbf{Exponential decay zone} (01:30:54-01:35:23 UT)} & \multicolumn{2}{c|}{\textbf{Power-law decay zone} (00:48:23-01:22:03 UT)} \\
\textbf{} & \multicolumn{4}{c||}{$\overline{V}_A=62kms^{-1}, \overline{T}_p=16.5eV, \overline{\beta}=0.8, \overline{B}_0=8nT$} & \multicolumn{2}{c|}{$\overline{V}_A=50kms^{-1}, \overline{T}_p=17.5eV, \overline{\beta}=1.1, \overline{B}_0=6nT$}
\\
\textbf{} & \multicolumn{4}{c||}{ $L_{zone}=1.96\times10^{5}km$, $M_{A,turb}=0.22$} & \multicolumn{2}{c|}{$L_{zone}=1.46\times10^{6}km$, $M_{A,turb}=0.31$}\\

\hline
Energy (keV) & $\Delta t_d$ (s) & $\kappa_{rr}$ ($km^2s^{-1}$) & $\kappa_{\parallel}$ ($km^2s^{-1}$) & $\lambda_\parallel (km)$ & Power-law index $\gamma$ & Diffusion exponent $\alpha$ \\
\hline
27.7 - 75.3
& 107.3 & $2.09\times10^7$ & $7.05\times10^7$ & $6.74\times10^4$ & 0.72 & 1.28 \\
92.2 - 160 & 256.1 & $5.00\times10^7$ &$1.68\times10^8$ & $1.03\times10^5$ & 0.85 & 1.15 \\
160 - 374 & 477.5 &  $9.32\times10^7$ & $3.14\times10^8$ &$1.32\times10^5$ & 0.80 & 1.20 \\
374 - 962 & 1929.8 &  $3.77\times10^8$ & $1.27\times10^9$ & $3.37\times10^5$ & 0.54 & 1.46 \\
\hline
\end{tabular}
\begin{tablenotes}
  \item $\overline{V}_A$ is the mean Alfv\'en speed, $\overline{T}_p$ is the mean proton temperature, $\overline{\beta}$ is the mean ratio of the plasma to magnetic pressure, $\overline{B}_0$ is the mean magnetic field, and $M_{A,turb}$ is the turbulent Alfv\'en Mach number. $L_{zone}$ represents the length of the exponential (power-law) decay zone, calculated as the product of the shock speed and the duration of each respective zone. 
\end{tablenotes}
\end{table*}

\begin{table*}[ht]
\label{table3}
\tablenum{3}
\caption{Diffusion Parameters of Energetic Electrons Upstream of the IP Shock on 2004 January 22}
\centering
\begin{tabular}{|c||c|c||c|c|}
\hline
\textbf{Electron} & \multicolumn{2}{c||}{\textbf{Near the IP shock}} & \multicolumn{2}{c|}{\textbf{Further from the IP shock}} \\
\hline
Energy (keV) & Power-law index $\gamma$ & Diffusion exponent $\alpha$ & Power-law index $\gamma$ & Diffusion exponent $\alpha$ \\
\hline
39.2 - 50.5 & 0.13 & 1.87 & 0.32 & 1.68 \\
50.5 - 68.1 & 0.16 & 1.84 & 0.34 & 1.66 \\
68.1 - 94.5 & 0.16 & 1.84 & 0.34 & 1.66 \\
94.5 - 128  & 0.15 & 1.85 & 0.36 & 1.64  \\
128 - 244   & 0.13 & 1.87 & 0.37 & 1.63 \\
244 - 406.5 & 0.12 & 1.88 & 0.40 & 1.60 \\
\hline
\end{tabular}
\end{table*}

\begin{table*}[ht]
\label{table4}
\tablenum{4}
\caption{First-order ($n=1$) gyroresonance Parameters Excited by Energetic Protons}
\centering
\begin{tabular}{|c||c|c|c|c|c|}
\hline
$i$&$E_i$ (keV) & $v_i$  $(km s^{-1})$ & \textbf{$k_{i\parallel,min}$ $(km^{-1})$} & $r_{ci}$ $(km)$ & $\Gamma_{gr}$ $(s^{-1})$ \\ \hline 
1&75.3 & $3.8\times10^3$ & $1.7\times10^{-4}$ & $7.0\times10^3$ & $0.38A_2-0.51A_2$\\ 
2&92.2 & $4.2\times10^3$ & $1.5\times10^{-4}$ & $7.7\times10^3$ & $0.30A_2-0.37A_2$\\ 
3&160  & $5.5\times10^3$ & $1.1\times10^{-4}$ & $1.0\times10^4$ & $0.17A_2$\\ 
4&374  & $8.5\times10^3$ & $7.4\times10^{-5}$ & $1.6\times10^4$ & $0.06A_2-0.08A_2$\\ 
5&962  & $1.4\times10^4$ & $4.5\times10^{-5}$ & $2.6\times10^4$ & $0.02A_2-0.03A_2$\\ 
\hline
\end{tabular}
 \begin{tablenotes}
 \item $*A_2$ is the second-order pitch angle anisotropy, which is presented in Figure \ref{figure3}(g). 
\end{tablenotes}
\end{table*}


\begin{figure*}
\centering
\includegraphics[scale=0.6]{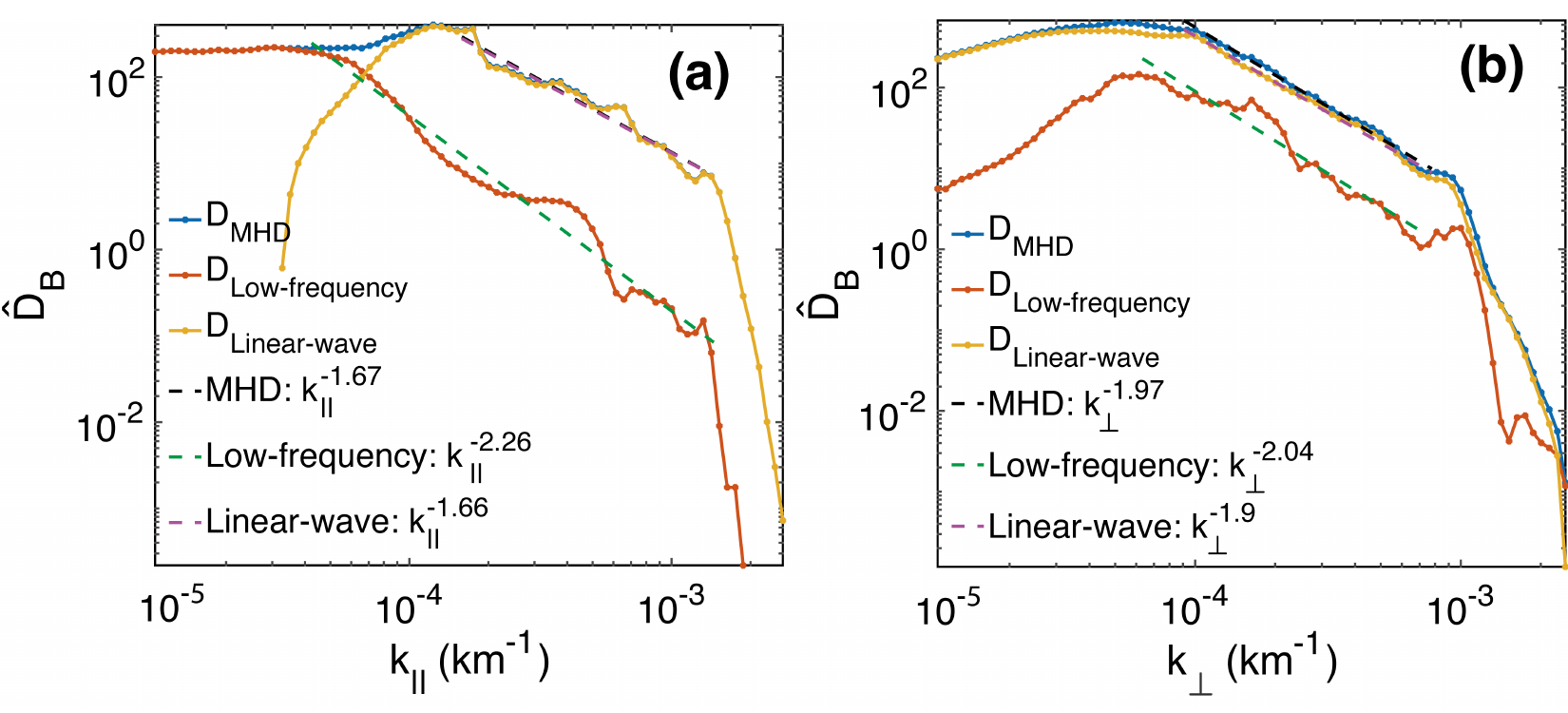}
\caption{Reduced energy density of magnetic field fluctuations in the power-law decay zone. (a) Normalized parallel wavenumber distribution of magnetic energy ($\hat{D}_B(k_\parallel)$). (b) Normalized perpendicular wavenumber distribution of magnetic energy ($\hat{D}_B(k_\perp)$). For convenience of comparison, all magnetic energy spectra are normalized by $B_0^2$ in the power-law decay zone. Blue curves represent all fluctuations on MHD scales, red curves represent the nonlinear low-frequency component of MHD-scale fluctuations, and yellow curves represent the linear wave component of MHD-scale fluctuations. The dashed lines denote the power-law fits.}
\label{figure8}
\end{figure*}

\begin{figure*}
\centering
\includegraphics[scale=0.33]{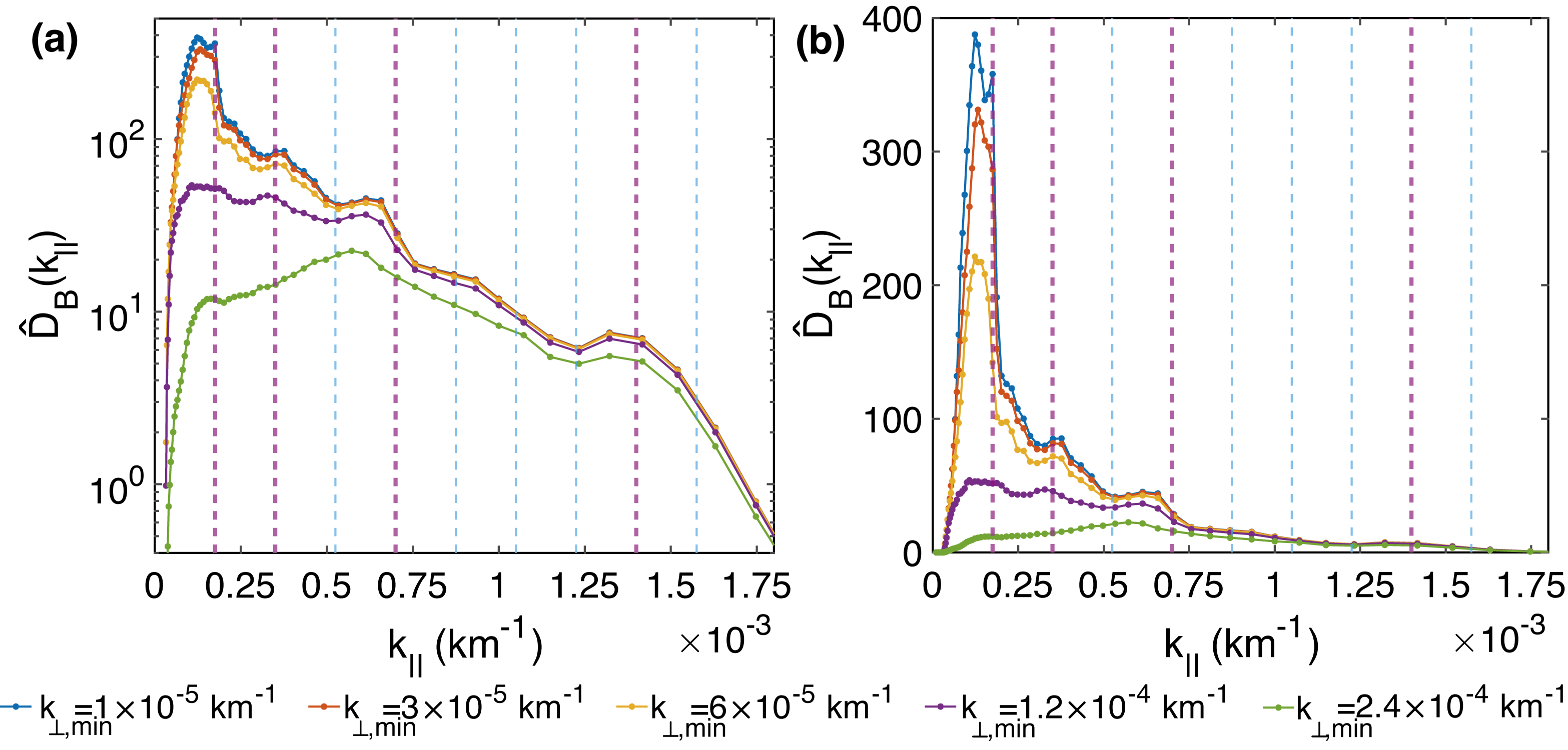}
\caption{Linear wave component of MHD-scale fluctuations: Normalized wavenumber distributions of magnetic energy $\hat{D}_B(k_\parallel)$ in log-linear (a) and linear-linear (b) scales. The pink dashed lines mark $k^*$, $2k^*$, $4k^*$, and $8k^*$, whereas the light blue dashed lines mark $3k^*$, $5k^*$, $6k^*$, $7k^*$, and $9k^*$, where $k^*=1.75\times10^{-4}km^{-1}$. For convenience of comparison, all magnetic energy spectra are normalized by $B_0^2$ in the power-law decay zone.}
\label{figure9}
\end{figure*}

\begin{figure*}
\centering
\includegraphics[scale=0.6]{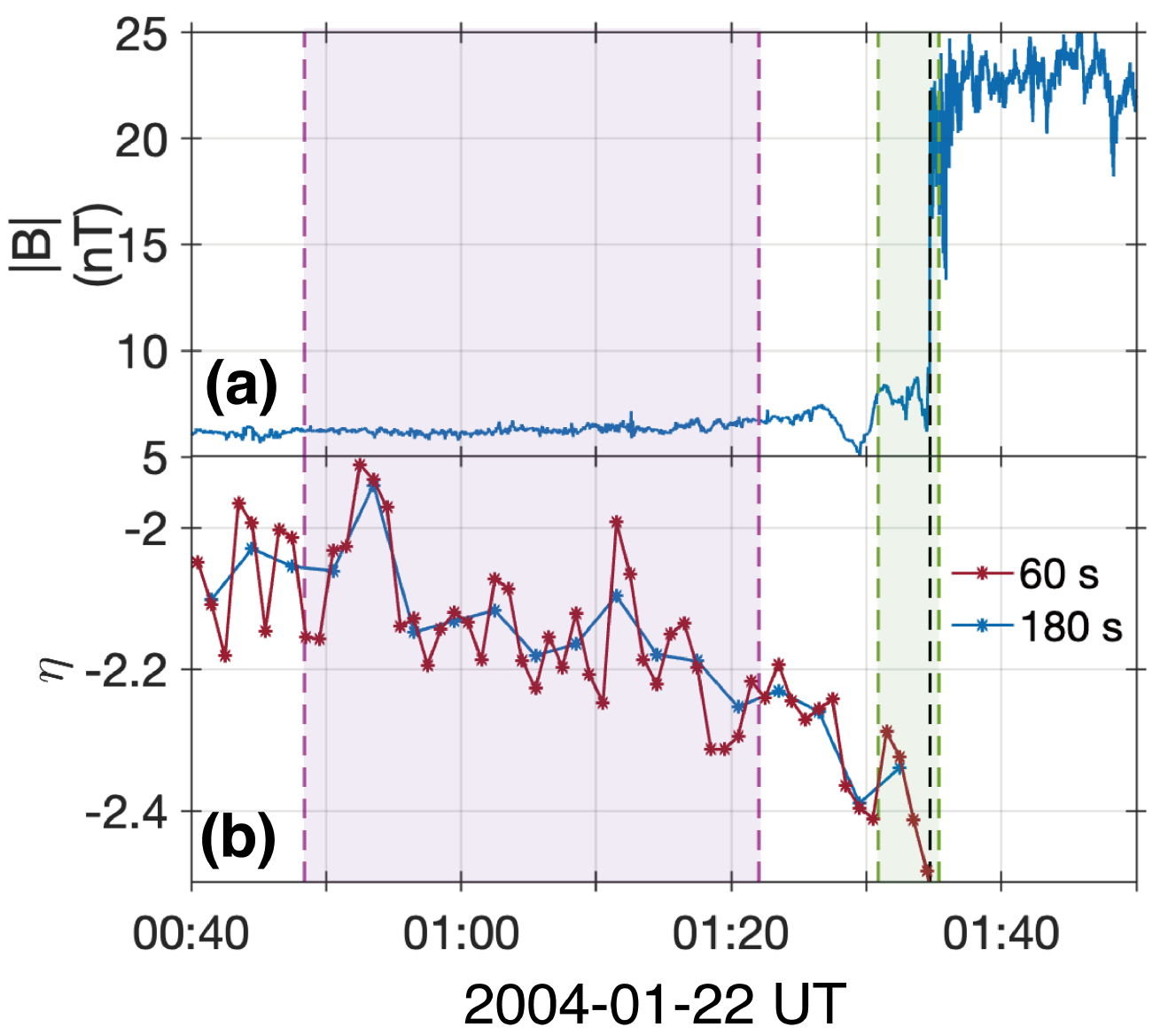}
\caption{Variations in the energy spectra of energetic protons upstream of the IP shock observed on 2004 January 22. (a) Magnetic field. (b) Power-law indices ($\eta$) of $dN/dE$ versus energy ($E$), fitted over the energy range [75.3, 962] keV and calculated using moving windows of 60 s and 180 s. The purple (green) shaded regions mark the power-law (exponential) decay zone for energetic protons.}
\label{figure10}
\end{figure*}

\bibliography{sample631}{}
\bibliographystyle{aasjournal}

\end{document}